# Spin wave mode coexistence on the nano-scale: A consequence of the Oersted field induced asymmetric energy landscape

Randy K. Dumas[1,*], E. Iacocca[1], S. Bonetti[2], S.R. Sani[3,4], S.M. Mohseni[3,4], A. Eklund[5], J. Persson[4], O. Heinonen[6,7], and Johan Åkerman[1,3,4]

[1]Physics Department, University of Gothenburg, 412 96 Gothenburg, Sweden

[2]Stanford Institute for Energy and Materials Science, Stanford University, Stanford, CA, USA

[3]Materials Physics, School of ICT, Royal Institute of Technology (KTH), 164 40 Kista, Sweden

[4]NanOsc AB, 164 40 Kista, Sweden

[5]Devices and Circuits, School of ICT, Royal Institute of Technology (KTH), 164 40 Kista, Sweden

[6]Materials Science Division, Argonne National Laboratory, Lemont, IL 60439 USA

[7]Department of Physics and Astronomy, Northwestern University, Evanston IL 60208, USA

It has been argued that if multiple spin wave modes are competing for the same centrally located energy source, as in a nanocontact spin torque oscillator, that only one mode should survive in the steady state. Here, the experimental conditions necessary for mode *coexistence* are explored. Mode coexistence is facilitated by the local field asymmetries induced by the spatially inhomogeneous Oersted field, which leads to a physical separation of the modes, and is further promoted by spin wave localization at reduced applied field angles. Finally, both simulation and experiment reveal a low frequency signal consistent with the intermodulation of two coexistent modes.

*Corresponding Author: randydumas@gmail.com

Interest in spin wave (SW) based electronics, also known as magnonics [1, 2], has resulted in increased research geared towards the reliable generation, manipulation, and detection of SWs. As opposed to cumbersome inductive techniques, spin transfer torque [3-5] offers an efficient means to convert a dc current into GHz magnetization oscillations, and is therefore ideally suited as a local and intrinsically nanoscale SW injector [6-8]. Nanocontact based spin torque oscillators (NC-STOs) [9-11] offer an ideal platform to study SW generation, owing to their extended free layer which potentially allows for unencumbered SW propagation. A popular school of thought regarding multi-mode generation centers on the fact that if each mode is competing for the same centrally located energy source, provided by the spin transfer torque of a single NC, only one mode should survive in the steady state [12]. This "survival of the fittest" approach is in analogy to two species competing for the same food supply. However, as evidenced in both experiments [13-19] and simulations [13, 15] clear signatures of mode-hopping are most often observed. In this case, while the system is free to explore two or more modes; at any given instant the system can only be found in a single mode. Here, we present a third possibility, namely that under the correct experimental conditions true mode *coexistence* can be realized in NC-STOs. Mode coexistence is facilitated by the local field asymmetries induced by the spatially inhomogeneous Oersted field in the vicinity of the NC and further promoted by mode localization at reduced field angles.

The seminal work of Slonczewski [20] predicted exchange dominated propagating SWs when the free layer is magnetized perpendicularly. The subsequent experimental work of Bonetti *et al*. [15] and Madami *et al.* [6] then clearly demonstrated the Sloncszewski-like propagating mode. Interestingly, the experimentally observed dynamical processes were shown to vary dramatically when the free layer is magnetized in-plane [10]. Furthermore, the behavior becomes significantly more complex as the free layer magnetization is biased at intermediate oblique angles. To summarize, while the linear propagating Slonczewski mode is stable for all applied field angles, only below a certain critical applied field angle, $\theta < \theta_C$, does a second fundamentally different type of SW excitation also become energetically favorable. This second mode is classified a highly non-linear localized solitonic SW bullet, as found both analytically [21, 22] and in micromagnetic simulations [23]. Simulations [15] and experiments [14] then provided evidence of mode-hopping between the propagating and localized bullet modes for $\theta<\theta_C$.

Samples for this study are based on NC-STOs. A circular NC of diameter $d$ is defined through a $SiO_2$ insulating layer using $e$-beam lithography on top of a 16 $\mu$m x 8 $\mu$m spin valve mesa. The magnetically active portion of the mesa is composed of a (nominal thicknesses in nm) $Co(8)/Cu(8)/Ni_{80}Fe_{20}(4.5)$ pseudo spin valve stack deposited by magnetron sputtering. The $Ni_{80}Fe_{20}$ and Co play the role of the *free* and *fixed* layers, respectively All measurements were performed at room temperatures utilizing in a highly uniform and precisely rotatable field with a fixed magnitude of $\mu_\circ H$= 0.965 T produced by a Halbach array of permanent magnets; further details of the electrical setup can be found in Ref. [24]. Finally, microwave excitations were only observed for a negative current polarity corresponding to electron flow from the free to the fixed magnetic layer.

Micromagnetic simulations were carried out using the MuMax2 code [25]. The simulation volume comprises only the free $Ni_{80}Fe_{20}$ layer and is a disk with a diameter of 1000 nm with a cell size of 3.9 x 3.9 x 4.5 $nm^3$ and highly absorbing ($\alpha$=1) boundaries to minimize SW reflections. A uniform spin-polarized current then acts on a sub-volume of the free layer with diameter *d*, matching the nominal experimental NC diameter. The current-induced Oersted field is also included in the simulations assuming current flow along an infinite cylinder [26]. The material parameters used for the $Ni_{80}Fe_{20}$ are a saturation magnetization $\mu_\circ M_S$=0.88 T, as determined by FMR measurements, exchange constant $A$=1x$10^{-11}$ J/m, anisotropy $K$=0 J/m$^3$, and dimensionless damping parameter α=0.01. The internal magnetization angle, which sets the spin polarization angle, of the Co fixed layer ($\mu_\circ M_S$=1.7 T) is evaluated by solving the magnetostatic boundary conditions in an external field of $\mu_\circ H$=0.965 T applied at an angle $\theta$ with respect to the film plane. A spin torque efficiency of 0.3 is used and provides excellent quantitative agreement with the experimentally observed frequencies. Finally, no interlayer exchange coupling between the fixed and free layers is taken into account and the simulations were performed at $T$=300 K following Brown's thermal field formulation [27]. Micromagnetic simulations that explicitly take into account the magnetization dynamics in both the fixed and free layers have also been carried over a smaller parameter set and are in agreement with those presented here [28].

The angular dependence of the generated microwave signals at a constant current of $I_{dc}$=-20 mA for a device with a nominal NC diameter of $d$=90 nm is shown in Fig. 1(a). The general behavior is consistent with a prior report on a much smaller NC diameter of $d$= 40 nm [15]. Above the critical angle, $\theta_C$≈60º, only a single mode with an oscillation frequency above the ferromagnetic resonance (FMR) frequency of the extended $Ni_{80}Fe_{20}$ film is observed and has been attributed to a propagating SW [6, 15] in the $Ni_{80}Fe_{20}$ free layer. However, for applied field angles less than $\theta_C$ a second mode with a frequency far below the FMR frequency, corresponding to the localized solitonic SW bullet, also exists [15, 21]. The angular dependence of the peak power and linewidth are shown in Fig. 1(b) and 1(c), respectively. The dramatic drop in power of the propagating mode for $\theta<\theta_C$ is expected given the total available energy now has to be shared between two modes, while the increase in linewidth has been attributed to mode-hopping [14]. Observed within the angular range 40°<$\theta$<$\theta_C$ is a broad low frequency ($f$<~2 GHz) signal that can also be attributed to mode-hopping, and will be discussed in more detail later. As the applied field angle is reduced below 45° the linewidth shows an order of magnitude decrease accompanied by a small increase in peak power for each mode. Most interestingly, this decrease (increase) in linewidth (power) is coincident with the applied field angle at which the frequency of the propagating mode becomes lower than the FMR frequency of the $Ni_{80}Fe_{20}$ free layer, suggesting that the localization of the propagating mode is playing an important role in improving the quality factor of each mode. Even though the higher frequency propagating mode drops below the FMR frequency of the extended $Ni_{80}Fe_{20}$ film for low angles, and will become highly localized in nature, we will continue to refer to this highest frequency mode as the propagating mode in order to stress that this is the same mode that propagates at higher angles. Finally, for $\theta$<20° the power of each mode decreases below the point of experimental detection as the precession angle of each mode becomes vanishingly small [14].

Micromagnetic simulations for $I_{dc}$=-20 mA and a NC diameter of $d$=90 nm, Fig. 1(d), quantitatively reproduce all of the experimentally observed behavior, including the broad low frequency signal for 40°<$\theta$<$\theta_C$. To shed more light on the dynamics in this angular range the simulated time evolution of the normalized $M_x$ component averaged under the NC at $\theta$=45° is shown in Fig. 1(e). The simulated behavior is very erratic and the $M_x$ component randomly switches from being predominately parallel to anti-parallel with respect to the in-plane component of the external bias field, which is directed along the $+x$ direction. Similarly, the smoothed pseudo-Wigner-Ville distribution of the device resistance, Fig. 1(f), exhibits a very complex and seemingly random evolution of the generated frequencies. Prior simulations revealed that the trajectory of the localized mode lies primarily anti-parallel to the external field direction [14, 29]. This erratic behavior is then consistent with a random mode-hopping between the propagating and localized modes which predominately oscillate parallel and anti-parallel to the external bias field direction, respectively. Additionally, we can now understand the origin of the broad low frequency signal as arising from the relatively slow and highly stochastic mode-hopping occurring for this range of applied field angles. Also evident in the simulated spectra for $20° < \theta < \sim 45°$ is a weak and very narrow low frequency signal that occurs at exactly the difference between the localized and propagating modes, as shown in Fig. 2 for $\theta=30°$. This is consistent with the 2$^{nd}$ order intermodulation signal of the localized ($f_L$) and propagating ($f_P$) modes occurring at $f_P$-$f_L$. This intermodulation signal is not apparent in the experimentally observed spectra, Fig. 1(a), but may be below our experimental detection limit for this NC size. Simulations with varying NC diameters, and a constant current density of $J=5.19 \times 10^{12}$ A/m$^2$, reveal that this intermodulation signal not only always occurs exactly at the difference between the localized and propagating mode frequencies, but also becomes significantly more powerful as the NC diameter is decreased, Fig. 2. Simulations for the smallest NC diameter, $d=60$ nm, also show a 3$^{rd}$ order intermodulation signal at $2f_L$-$f_P$. In fact, the 2$^{nd}$ order intermodulation difference signal is routinely measured for a smaller NC diameter of $d=70$ nm, Fig. 3(a). Fig. 3(a) shows the spectra measured as a function of $I_{dc}$ at a fixed applied field angle of $\theta=30°$ where both the propagating and localized modes are excited. Consistent with a prior report [15] the lower (higher) frequency localized (propagating) mode shows a clear decrease (increase) in frequency as the magnitude of $I_{dc}$ is increased. Correspondingly, the 2$^{nd}$ order intermodulation signal shows a steady increase in frequency. Micromagnetic simulations, Fig. 3(b), again quantitatively reproduce all of the experimentally observed features including the threshold current, intermodulation signal, and the current tunability ($df/dI_{dc}$) of each mode. In stark contrast to the behavior observed in Fig. 1(e), the simulated time evolution of the normalized $M_x$ component, Fig. 3(c), shows a much more regular, periodic, and predictable behavior at an applied field angle of $\theta=30º$. Furthermore, unlike the behavior exhibited at $\theta=45º$, Fig. 1(e), the oscillation trajectory is now primarily opposing the external bias field direction. The characteristic beating pattern shown in Fig. 3(c) is consistent with the simple summation of two coexistent frequencies, as further supported by the smoothed pseudo-Wigner-Ville distribution, shown in Fig. 3(d). Cross-terms, circled in Fig. 3(d), are artifacts arising from the Wigner-Ville transform which actually provide additional information about the relative phase between the two frequencies. The observed behavior reveals that the relative phase relation between the two modes is changing in a well-defined and periodic manner as expected for two coexistent modes.

The spatial distribution of the modes provides additional insights into the observed mode coexistence and role played by the spatially inhomogeneous Oersted field [30, 31]. By separately evaluating the fast Fourier transforms of each simulation cell and filtering each image around the localized, $f_L$, or propagating, $f_P$, mode frequencies the spatial profiles of each mode can be mapped. The spatial distribution of the propagating and localized modes for $I_{dc}=-18$ mA and $\theta=30º$ are shown in Fig. 4(a) and 4(b), respectively, highlighting that the modes are spatially separated [32]. The localized mode power is primarily confined to an arc around the upper half of the NC diameter, whereas the propagating mode power is centered at a point where the in-plane component of the external field, $H_{//}$, and Oersted field produces a local field maximum at the bottom of the NC. At $\theta=30º$ each mode lies significantly below the FMR frequency of the surrounding $Ni_{80}Fe_{20}$ film and therefore the energy density of each mode is highly localized to regions very near the NC. This is in contrast to a higher angles were the frequency of the propagating mode lies well above the FMR frequency [28]. It is also interesting that the majority of the power for each mode exists in a region just outside the NC diameter where the current density is zero. This can be simply understood by considering the fact that for the region inside the NC the average *x*-component of the magnetization primarily points to the left, i.e. anti-parallel to $H_{//}$ and schematically represented in Fig. 4(a). However, for the entirety of the region outside of the NC the *x*-component of the magnetization aligns parallel to $H_{//}$. In order to connect these two regions with opposing magnetization directions the largest angle precessions, and therefore maximum mode power, must occur just outside the boundary of the NC.

Experimental evidence for the physical separation of the modes can be found in the observed current tunability ($df/d|I_{dc}|$), Fig. 3(a) and 3(b). Prior analytical calculations [21], that do not take into account the spatially inhomogeneous Oersted field, predicted that current tunability of the localized and propagating modes would both "redshift", that is $df/d|I_{dc}|<0$, for applied field angles less than $\theta_C$. Additionally, micromagnetic simulations where the Oersted field is neglected confirmed this behavior [23]. However, experiments invariably have found that that the lower frequency localized mode redshifts, as expected, but that the high frequency propagating mode actually "blueshifts" ($df_P/d|I_{dc}|>0$) for all applied field angles. The changing local magnetic field environment, that is how $dH/d|I_{dc}|$ varies due to the changing current induced Oersted field, plays a significant role in determining the final current tunability of the propagating mode. A simple calculation assuming the Oersted field of an infinite wire with a diameter of 70 nm results in local changing field environment of $dH/d|I_{dc}|=+56$ Oe/mA on the lower boundary of the NC, where the propagating mode is centered. At a constant $I_{dc}$ the frequency tunability with externally applied field has been measured as $df_P/dH=2.7$ MHz/Oe. Therefore, based solely on the varying local field environment at the lower edge of the NC we would expect a current tunability of $df_P/d|I_{dc}|=+0.15$ GHz/mA for the propagating mode, which is

nearly double the experimentally observed and simulated tunability of $df_P/d|I_{dc}|=+0.07$ GHz/mA. Therefore, the changing local field environment induced by the Oersted field at the bottom edge of the NC is responsible for flipping the sign of the propagating mode tunability from negative, as expected from analytical calculations, to positive, as observed experimentally. The physical separation of the modes can also explain the general behavior highlighted in Fig. 2, where it was found that the strength of the intermodulation signal increased as the diameter of the NC decreased. As the NC diameter is decreased, there is more overlap and therefore more potential for non-linear mixing between the two modes. This results in a more prominent intermodulation signal for smaller NC diameters, consistent with what is also observed experimentally.

In conclusion, evidence of SW mode coexistence on the nanoscale was presented in NC-STOs. The spatially inhomogeneous Oersted field in the vicinity of the NC plays two important roles. Firstly, the broken symmetry of the local fields promotes a physical separation of the localized and propagating modes. Experimental evidence of the mode separation lays the observed current tunability of the propagating mode. Secondly, localization of the propagating mode for reduced field angles was found. Together, spatial separation and localization promotes coexistence and results in a distinct intermodulation signal, as also observed experimentally.

**Acknowledgements**

Support from The Swedish Research Council (VR), The Swedish Foundation for Strategic Research (SSF), and the Knut and Alice Wallenberg Foundation is gratefully acknowledged. Argonne National Laboratory is a US DOE Science Laboratory operated under contract no. DE-AC02-06CH11357 by UChicago Argonne, LLC.

**Figure Captions**

**FIG. 1** (color online). (a) Experimentally measured frequencies with corresponding (b) peak power and (c) linewidth and the (d) simulated frequencies as a function of the applied field angle for a NC-STO with a nominal NC diameter $d$=90 nm at a measurement current of $I_{dc}$=-20 mA. (e) Time evolution of the $M_x/M_S$ component of the magnetization and (f) smoothed pseudo-Wigner-Ville distribution of the simulated device resistance averaged under the NC for $\theta$=50°. The smoothed pseudo-Wigner-Ville distribution is evaluated using a frequency window of 2 GHz and a time window of 0.1 ns.

**FIG 2.** (color online). Simulated frequency spectra for several NC diameters at a constant current density of $J$=5.19x$10^{12}$ A/m$^2$ and applied field angle of $\theta$=30°.

**FIG. 3** (color online). (a) Experimental and (b) simulated frequencies as a function of $I_{dc}$ for a NC-STO with a nominal NC diameter of $d$=70 nm and applied field angle of $\theta$=30°. (c) Time evolution of the $M_x/M_S$ component of the magnetization and (d) smoothed pseudo-Wigner-Ville distribution of the simulated device resistance averaged under the NC for $I_{dc}$=-18 mA. The smoothed pseudo-Wigner-Ville distribution is evaluated using a frequency window of 0.4 GHz and a time window of 0.2 ns.

**FIG. 4** (color online). Simulated spatial profiles of the (a) propagating and (b) localized modes for a NC-STO with a NC diameter of $d$=70 nm and applied field angle of $\theta$=30° and $I_{dc}$=-18 mA. The yellow solid circle defines the NC diameter.

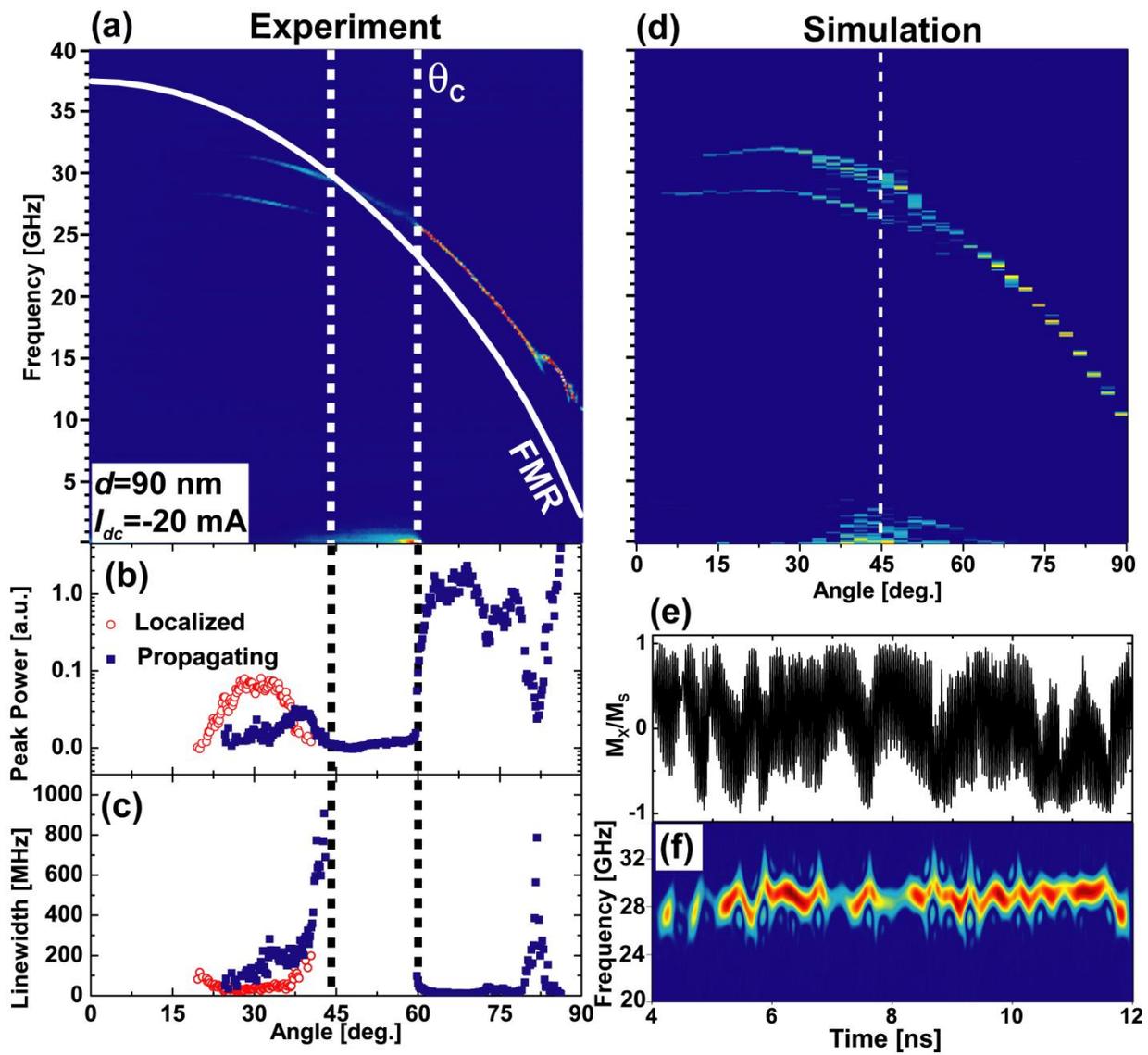

**Fig. 1, Dumas *et al.***

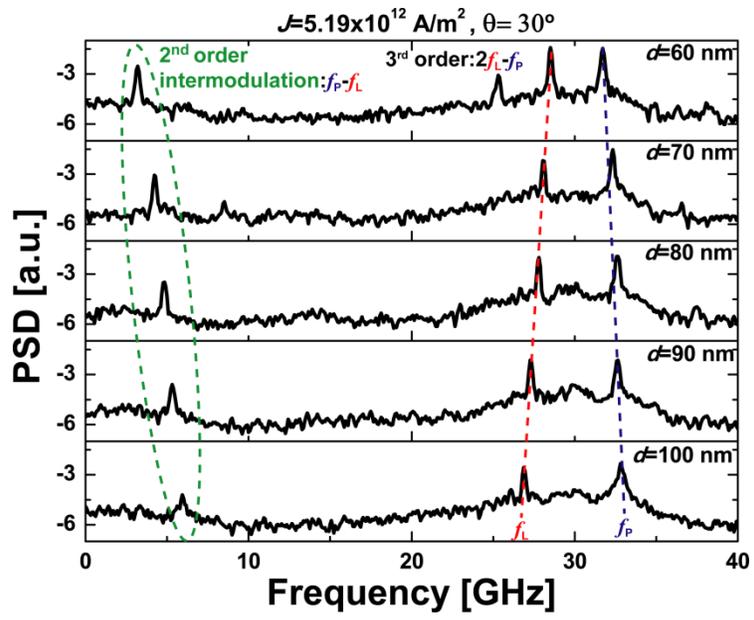

**Fig. 2,** Dumas *et al.*

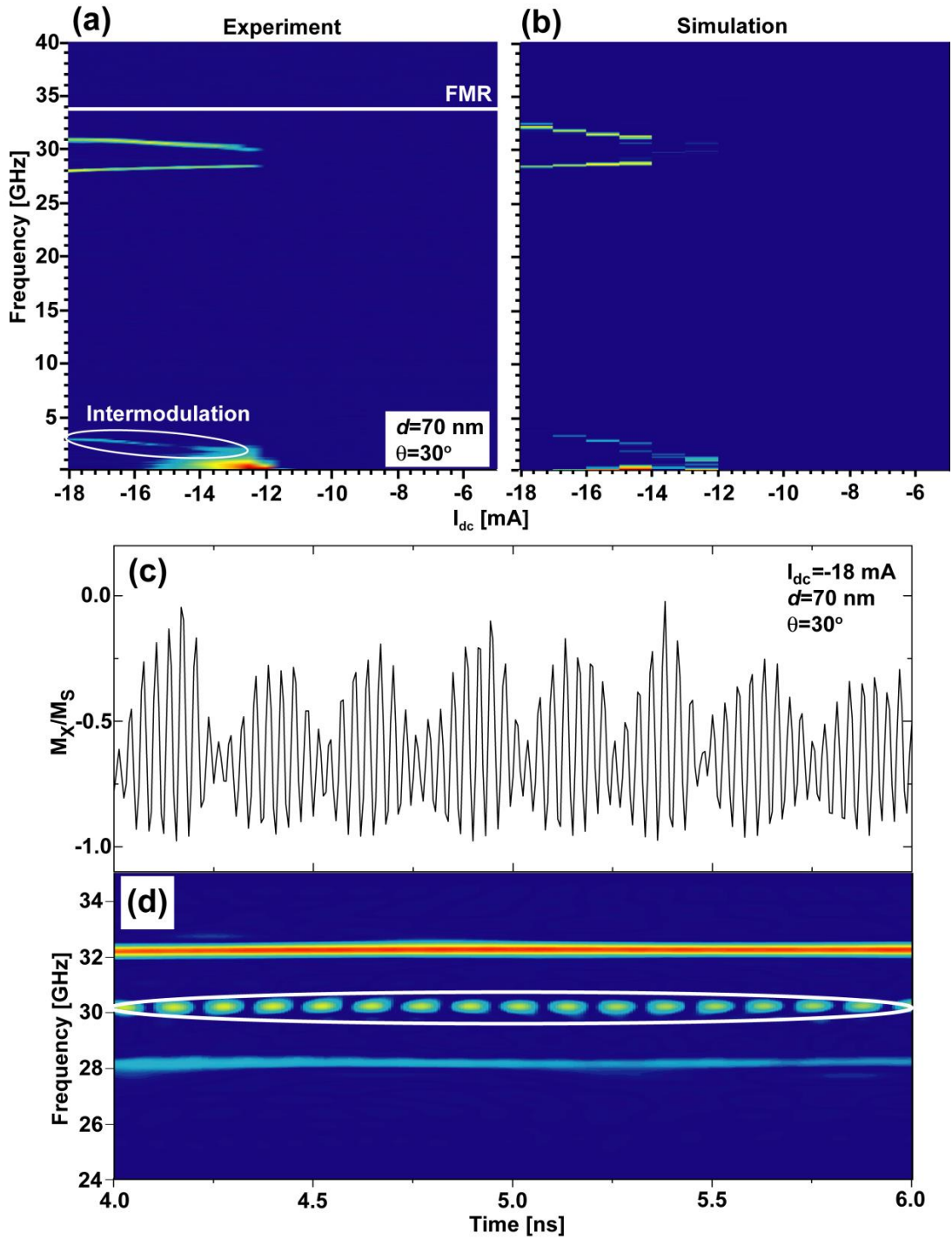

**Fig. 3, Dumas *et al.***

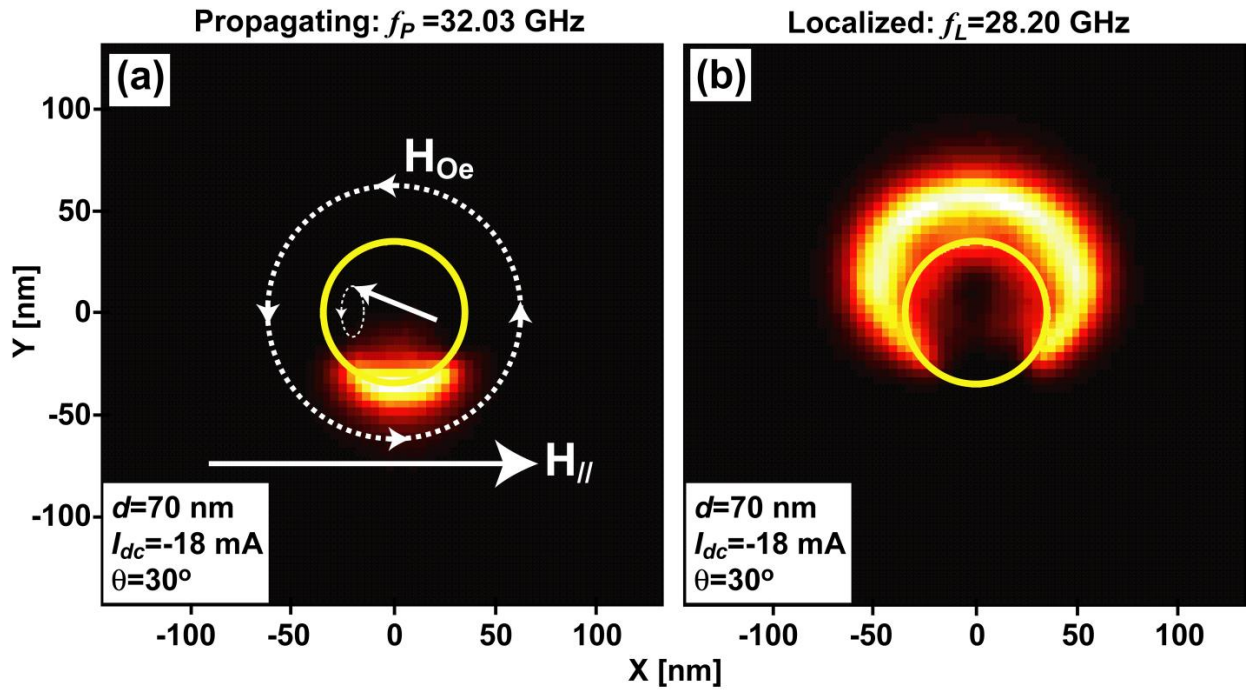

**Fig. 4, Dumas *et al.***

# Supplementary Material

# Spin wave mode coexistence on the nano-scale: A consequence of the Oersted field induced asymmetric energy landscape

Randy K. Dumas[1], E. Iacocca[1], S. Bonetti[2], S.R. Sani[3,4], S.M. Mohseni[3,4], A. Eklund[5], J. Persson[4], O. Heinonen[6,7], and Johan Åkerman[1,3,4]

[1]Physics Department, University of Gothenburg, 412 96 Gothenburg, Sweden

[2]Stanford Institute for Energy and Materials Science, Stanford University, Stanford, CA, USA

[3]Materials Physics, School of ICT, Royal Institute of Technology (KTH), 164 40 Kista, Sweden

[4]NanOsc AB, 164 40 Kista, Sweden

[5]Devices and Circuits, School of ICT, Royal Institute of Technology (KTH), 164 40 Kista, Sweden

[6]Materials Science Division, Argonne National Laboratory, Lemont, IL 60439 USA

[7]Department of Physics and Astronomy, Northwestern University, Evanston IL 60208, USA

**Micromagnetic Simulations**

In addition to the MuMax2 simulations presented in the main text, micromagnetic simulations that take into account the dynamical processes in **both** the fixed and free layers were also carried out and are presented here. The simulation volume comprises the $Ni_{80}Fe_{20}$ free layer, Cu spacer, and Co fixed layer. A disk with a diameter of 500 nm is used, where highly absorbing boundaries ($\alpha=1$) are implemented to minimize spin wave reflections. The cell size has surface dimensions of 5x5 nm$^2$ while the thickness is adjusted to each layer. A uniform spin-polarized current then acts on a sub-volume of the free layer with diameter $d=70$ nm, matching the nominal experimental nanocontact diameter. The current-induced Oersted field is also included in the simulations assuming current flow along an infinite cylinder and an external field of $\mu_\circ H=0.965$ T is applied at an angle $\theta=30º$ with respect to the film plane. The material parameters used for the $Ni_{80}Fe_{20}$ layer are a saturation magnetization $\mu_\circ M_S=1.0$ T, exchange constant $A=1$x$10^{-11}$ J/m, and damping $\alpha=0.01$. The material parameters used for the Co layer are a saturation magnetization $\mu_\circ M_S=1.7$ T and exchange constant $A=1.2$x$10^{-11}$ J/m. Both interlayer interactions and crystalline anisotropy, as the films are polycrystalline, are neglected. A spin polarization of 0.65 (equivalent to the spin torque efficiency used in the paper) is assumed for the Co fixed layer. The spin polarization of the $Ni_{80}Fe_{20}$ layer is considered negligible and it was thus set to 0.1. The simulations were performed at $T = 300$ K. We should note that these material parameters were chosen to agree with nominally accepted values typically found in the literature and not altered to provide agreement with experiment.

The experimental (solid symbols) and simulated (open symbols) frequencies as a function of the dc drive current ($I_{dc}$) are shown in Fig. I(a) and exhibit very good quantitative agreement. The simulated spatial profiles of the localized (top) and propagating (bottom) modes are shown in Fig. I(b) for three selected drive currents. As the magnitude of the drive current is increased the propagating mode moves progressively closer to the lower boundary of the NC where $H_{Oe}$ and $H_{//}$ generate a local field maximum whereas the localized mode is preferentially generated towards the upper half of the NC.

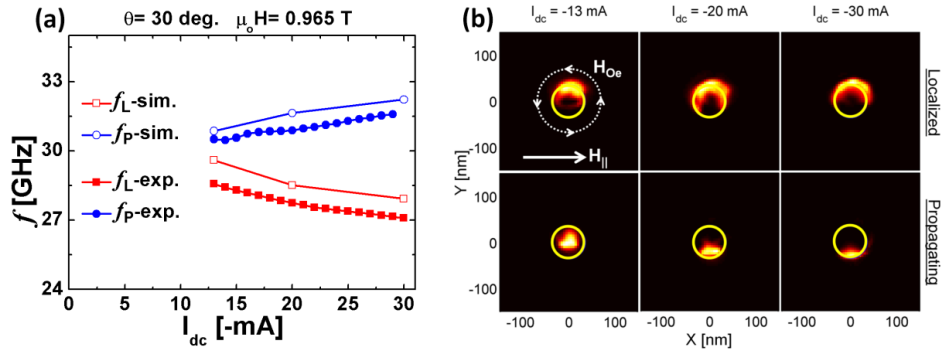

**Figure I.** Micromagnetic simulations. (a) Experimental (solid symbols) and simulated (open symbols) frequencies as a function of $I_{dc}$. (b) Simulated spatial profiles of the localized (top) and propagating (bottom) modes for three select drive currents.

**θ=75º: Propagating Spin Waves**

Here we present simulations, using MuMax2, at a large applied field angle of θ=75º where only the propagating mode is excited and lies above the FMR frequency of the $Ni_{80}Fe_{20}$. By separately evaluating the fast Fourier transform of each simulation cell and filtering each image around the propagating mode frequency, $f_P$, the spatial profiles can be mapped. The spatial distribution of the propagating modes for $d$=90 nm, $I_{dc}$=-20 mA and $\theta$=75° is shown in Fig. II(a). Contrary to the behavior presented in the main text, Fig. 4a and 4b, the majority of the propagating mode power is now centered directly below the NC and shows strong radiation into the region above the NC. This asymmetric radiation profile is due to the spatially asymmetric Oersted field modifying the local field environment, and therefore the FMR frequency, in the vicinity of the NC.[1] For regions below (above) the NC the Oersted field adds to (subtracts from) the in-plane component of the external field, $H_{//}$, therefore locally increasing (decreasing) the FMR frequency. The propagating mode frequency (red dotted line), FMR frequency in the absence of an Oersted field (blue dashed line, $I_{dc}$=0 mA), and FMR frequency including the influence of the Oersted field (black solid line, $I_{dc}$=-20 mA) are shown in Fig. II(b) for $x$=0 in the regions outside the NC diameter. For regions just below the nanocontact the propagating mode frequency is initially less than the local FMR frequency which significantly inhibits propagation in this direction.

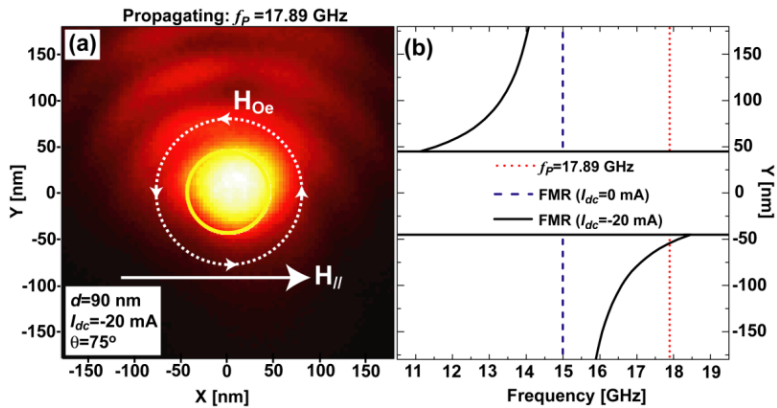

**Figure II.** Propagating spin waves. (a) Simulated spatial profile of the propagating mode for a NC-STO with a NC diameter of $d$=90 nm, applied field angle of $\theta$=30°, and $I_{dc}$=-20 mA. The yellow solid circle defines the NC diameter. (b) Calculated FMR frequency neglecting the Oersted field (blue dashed line) and including the current induced Oersted field (solid black line) are shown relation to the propagating mode frequency (red dotted line) along $x$=0.